\pdfoutput=1
\documentclass[aps,12pt,superscriptaddress,nofootinbib]{revtex4-2}

\usepackage[colorlinks=true,citecolor=blue,urlcolor=blue,linkcolor=blue,bookmarksopen]{hyperref}
\usepackage{amsmath,amssymb}
\usepackage{graphicx}
\usepackage{bm}
\usepackage{comment}
\usepackage{upgreek}

\usepackage{amsmath,amssymb}
\usepackage{bm}
\usepackage{graphicx}

\renewcommand\d{\partial}

\renewcommand\Im{\mathop{\mathrm{Im}}}

\renewcommand\d{\partial}
\newcommand\q{\mathbf{q}}
\newcommand\x{\mathbf{x}}
\newcommand\y{\mathbf{y}}
\renewcommand\k{\mathbf{k}}
\newcommand\J{\mathbf{J}}
\newcommand\B{\mathbf{B}}
\newcommand\+{\dagger}
\newcommand\<{\langle}
\renewcommand\>{\rangle}
\renewcommand\Im{\mathop{\mathrm{Im}}}
\newcommand\R{\mathbf{R}}

\begin{document}

\title{Probing the spin structure of the fractional quantum Hall magnetoroton with polarized Raman scattering} 
\author{Dung Xuan Nguyen}
\email{dung\_x\_nguyen@brown.edu}
\affiliation{Brown Theoretical Physics Center and Department of Physics, Brown University, 182 Hope Street, Providence, Rhode Island 02912, USA}
\author{Dam Thanh Son}
\affiliation{Kadanoff Center for Theoretical Physics, University of Chicago,
  933 East 56th Street, Illinois 60637, USA}
\date{\today}  
\begin{abstract}

Starting from the Luttinger model for the band structure of GaAs, we derive
an effective theory that describes the coupling of the fractional
quantum Hall (FQH) system with photons in resonant Raman scattering
experiments.  Our theory is applicable in the regime when the energy of the
photons
$\omega_0$ is close to the energy gap $E_G$, but $|\omega_0-E_G|$ is much
larger than the energy scales of the quantum Hall problem.
In the literature, it is often assumed that Raman scattering  measures
the dynamic structure factor
$S(\omega,\mathbf{k})$ of the FQH.  However, in this paper, we find
that the light scattering spectrum measured in the experiments are
proportional to the spectral densities of a pair of operators which we
identified with the spin-2 components of the kinetic part of the
stress tensor.
In contrast with the dynamic structure factor, these spectral densities
do not vanish in the long-wavelength limit $k\to0$.
We show that
Raman scattering with circularly polarized light can measure the spin
of the magnetoroton excitation in the FQH
system.  We give an explicit expression for the kinetic stress tensor
that works on any Landau level and which can be used for numerical calculations
of the spectral densities that enter the Raman scattering amplitudes.
We propose that Raman scattering provides a way to probe the bulk
of the $\nu=5/2$ quantum Hall state to determine its nature.

\end{abstract}

\maketitle

\section{Introduction}

The fractional quantum Hall effect (FQHE) was discovered in experiment
almost forty years ago \cite{Tsui}.  Fractional quantum Hall (FQH)
systems support a host of intriguing physical phenomenons; they are
also a playground for many exotic theoretical ideas ranging from
anyons~\cite{Laughlin:1983fy,Anyon2,Anyon1} to
superconductivity~\cite{ReadGreen}, skyrmion~\cite{Ezawa:SkyrmionFQH},
and bimetric gravity~\cite{Gromov2017}, to name a few.  Anyonic
excitations in a nonabelian FQH states such as the Moore-Read state at
filling fraction 5/2 \cite{MOOREREAD} may provide the building blocks
for a topological quantum computer
\cite{AnyonQuantumComp}.  However, FQHE is still one of the most
difficult and important unsolved problems of modern physics.

In a classic paper~\cite{Girvin1986}, Girvin, MacDonald, and Platzman
proposed a single mode approximation for the FQHE, in which the only
excitation of the FQH system is a gapped charge-neutral mode called
``magnetoroton.''  In the original treatment, the magnetoroton was
interpreted as the charge density wave in the lowest Landau level
(LLL).  The dispersion relation of this neutral mode has a minimum at
the wave length is of order the magnetic length $\ell_B$, which
imitate the behavior of the roton mode in superfluid $^4$He.
Experiments have confirmed the existence of the magnetoroton mode in
light scattering
experiments \cite{Pinczuk1,Pinczuk2,Kukushkin}~\footnote{For an
alternative interpretation of the mode, see
Ref.~\cite{Wiegmann:2018pru}.}.  In the widely accepted theoretical
interpretation of these experiments, the light scattering intensity
has been associated with the dynamical structure factor
$S(\omega,\mathbf{k})$ in the LLL~\cite{PlatzmanHe}.  However, in the
LLL limit, the dynamical structure factor goes to zero as
$k^4$ \cite{Girvin1986} in the limit where the momentum of the
magnetoroton $k$ goes to zero.  On the other hand, the Raman signal
seems to persist down to $k=0$, signaling that the identification of
the intenstity of Raman scattering with the dynamic structure factor
may not be correct.

In this paper, we provide a new theoretical treatment for Raman
scattering experiments.  We first note that the problem of Raman
scattering involves many energy scales.  The largest energy scale is
that of the semiconductor gap $E_G$ and the photon energy $\omega_0$.
The next scale is the distance between the Landau levels of the
conduction-band electrons, $\omega_c = eB/m^*c$, and the smallest
energy scale is the Coulomb interacting energy between these
electrons, $\Delta$.  We assume a hierarchy
\begin{equation}
  \Delta \ll \omega_c \ll E_G .
\end{equation}
Only the physics at the scale $\Delta$ is ``hard,'' i.e.,
nonperturbative or strongly correlated, while the physics at the
scales $E_G$ and and $\omega_c$ are weakly coupled.  To solve the
Raman scattering problem effectively, one needs to separate out the
nonperturbative, strongly correlated physics at the scale $\Delta$
from the perturbative, weak coupling physics at the other scales.
(This is similar to the philosophy of ``factorization'' in quantum
chromodynamics~\cite{Collins:1989gx} where the perturbative physics at
the hadronic scale is separated from the perturbative physics of
higher energy scales.)

We perform this ``factorization'' procedure in two steps.  The first
is to integrate out the energy scale $E_G$.  Starting with the
Luttinger's Hamiltonian for GaAs~\cite{Luttinger1956}, we introduce a
coupling between the lowest conduction band and highest valence band
due to the interaction with light waves.  We focus on the regime of
resonant Raman scattering in which the frequency of the incoming light
is close to the semiconductor energy gap: $\omega_0\approx E_G$.
Under the assumption that the detuning between the frequency of light
and the gap, $|\omega_0-E_G|$, is larger than both energy scales of
the Hall effect $\omega_c$ and $\Delta$, we integrate out valence
bands to obtain the coupling of the conduction-band electon to the
photon.  The second step is to do projection to a single Landau level.
The result is an effective coupling of the Raman photons to operators
acting on a single Landau level.

Our result differs drastically from that of Ref.~\cite{PlatzmanHe}.
We find that instead of measuring the spectral density of the density
operator (the dynamic structure factor), the Raman scattering
experiment measures the spectral densities of the operators that can
be called the ``kinetic stress tensor'' operators,
$T^{\text{kin}}_{ij}=\frac1{m^*}\d_i \psi^\+\d_j \psi$.  These
operatores are the components of the stress tensor that arise from the
kinetic energy term in the many-body Hamiltonian.  For simple model
Hamiltonians leading to exact zero-energy trial wave functions, the
kinetic stress tensor coincides with the full stress tensor, but that
is not the case for the general case, including that of Coulomb
interaction.  Moreover, in the lowest Landau level limit, in the
long-wavelength limit the only components of the kinetic stress tensor
that have nonvanishing spectral densities are the two spin-2
components, $T^{\text{kin}}_{zz}$ and $T^{\text{kin}}_{\bar z\bar z}$
(the spin-0 component, $T^{\text{kin}}_{z\bar z}$ has vanishing
spectral density in the limit $k\to0$).

Recent theoretical
works~\cite{Haldane:2011,Golkar,Liou2019,Son:2019qlm} advance a new
proposal on the interpretation of the magnetoroton.  According to this
proposal, the magnetoroton is the quantum a dynamical metric, and at
the long-wavelength regime $k\sim0$ has an angular momentum equals 2
in the direction of the applied magnetic field~\cite{Golkar}.  In
order to determine the spin of the magnetoroton, it has been
suggested~\cite{Golkar,Liou2019,Son:2019qlm} that the spin of the
magnetoroton can be detected through polarized Raman scattering.

We show it this paper how Raman scattering with cicrularly polarized
light can indeed be used to confirm the spin of the magnetoroton,
including the sign.  The basic idea is very simple: if angular
momentum is exactly conserved, an FQH state can absorb only one
specific circular polarized photon to excite a spin-2 magnetoroton
mode and emit a photon with opposite circular polarization.  However,
the real system does not have full rotational symmetry, but only
$C_4$, so spins 2 and $-2$ are not distinct from the point of view of
symmetry.  In this paper we carefully analyze the resonant Raman
scattering using the Luttinger model of GaAs.  We determine the
magnitude of the effect of nonconservation of angular momentum through
the Luttinger parameters and show that it is numerically small (of
order $1/40$).

We organize the paper as follows. We start Section~\ref{sec:model} by
introducing our theoretical model for Raman scattering of an FQH
state. In Section~\ref{sec:exp}, we present the calculation of the
intensity of cirularly polarized light scattering . We show that, in
contrast with previous theoretical proposals, the peaks in scattering
intensity, at the long wavelength regime, was obtained mainly due to
the poles in the correlation functions of the kinetic part of stress
tensor $\langle T^{\text{kin}}T^{\text{kin}}\rangle$.  We relate the
cross section of Raman scattering by circularly polarized light with
the spectral densities of the stress tensor. In Section~\ref{sec:TLL},
we derive the explicit formulas for the stress tensor projected to a
Landau level, which can be used to evaluate numerically the spectral
function.  We then conclude the paper in Section~\ref{sec:concl}.  The
Appendicies are devoted to the details of the calculation.

\section{Model}
\label{sec:model}

In GaAs the light hole and heavy hole bands make the main contribution
to Raman scattering process \cite{PlatzmanHe}. Thus the spin-off bands
can be ignored in this sense. We consider the effective Lagrangian,
with only conduction band $\psi^\alpha$ and $j=3/2$ valence bands
(which is nothing but light hole and heavy hole bands)
$\chi_i^\alpha$. In the notation, $\alpha=1,2$ represents spinor
index, and $i=1,2,3$ represents 3 components of $p$ wave function. The
(``Rarita-Schwinger'') constraint is imposed,
\begin{equation}
\label{eq:cont}
  {(\sigma^i)^\alpha}_\beta \chi^\beta_i=0,
\end{equation} 
which projects out the $j=1/2$ part from $\chi^\alpha_i$. Consider the
Luttinger Hamiltonian for heavy hole and light hole (within
approximation $\mathbf{k}\approx 0$)~\cite{Luttinger1956}
\begin{multline}\label{eq:Ham1}
  H = \frac1m \biggl\{
  \left(\gamma_1+\frac{5\gamma_2}2\right)\frac{\mathbf{D}^2}2-
  \gamma_2\sum_i  J_i^2  D_i^2 \\
  -2\gamma_3\left[\{ J_x J_y\}\{ D_x D_y\} + \{ J_y J_z\}\{D_y D_z\}
    + \{ J_z J_x\}\{D_z D_x\} \right]
  + \frac ec \kappa\J \cdot \B\biggr\},
\end{multline}
where $\gamma_{1,2,3}$ are the Luttinger's parameters, $m$ is the mass
of electron, $J_i$ are the SO(3) generators:
$(J_i)_{jk}=-i\epsilon_{ijk}$, $D_i\chi=(\d_i - i\frac ec A_i)\chi$ is
the covariant derivative,
$\{\cdot\}$ denotes symmetrization,
e.g., $\{D_x D_y\}\equiv \frac12(D_x D_y + D_y D_x)$, and $e<0$ is the
electric charge of the electron.  Using the equality
\begin{equation}
  \{ J_i J_j \} \{ D_i D_j \}
  = (\J \cdot \mathbf{D})^2 - \frac{e}{2c}\J\cdot\B ,
\end{equation}
we can rewrite Hamiltonian (\ref{eq:Ham1}) as
\begin{equation}
\label{eq:Ham11}
H = \frac{1}{m}\biggl[
  \Bigl(\gamma_1+\frac{5\gamma_2}{2}\Bigr)\frac{\mathbf{D}^2}{2}
  +(\gamma_3-\gamma_2)\sum_i  J_i^2 D_i^2
  -\gamma_3(\J\cdot\mathbf{D})^2
 + \frac{e}{c}\left(\kappa +\frac{\gamma_3}{2}\right)\J\cdot\B\biggr] .
\end{equation}

The Lagrangian for the hole band is 
\begin{equation}
  \mathcal{L}_v = i\chi^{\dagger}_{i\alpha}\partial_t\chi^\alpha_i
  -
  \chi^{\dagger}_{i\lambda}
  H_{ij}
  \chi^\lambda_j
  +E_G\chi^{\dagger}_{i\alpha}\chi^\alpha_i ,
\end{equation}
where $E_G$ is energy gap and the covariant derivative is
$D_i\psi \equiv (\partial_i-i\frac{e}{c}A_i)\psi$.
We also have Lagrangian of the conduction band
\begin{equation}
  \mathcal{L}_c=i\psi_\alpha^\dagger \partial_t \psi^\alpha
  -\frac{D_i\psi_\alpha^\dagger D_i\psi^\alpha}{2m^*} \,,
\end{equation}
and the coupling of the valence band and conduction band through
interaction with light
\begin{equation}
	\label{eq:photon}
  \mathcal{L}_{i} =
  e (P^*\psi_\alpha^\dagger \chi^\alpha_i E_i+P\chi^\dagger_{i\alpha}\psi^\alpha E_i) ,
\end{equation}
where $P$ is the strength of the dipole transition between the
conduction and valence bands (it will be related to the parameter
usually denoted as $E_p$ in the literature), $E_i$ is the electric
field of electromagnetic wave.

We consider the regime of resonant Raman scattering, where the photon
energy be is close to the gap: $|\omega_0 - E_G| \ll E_G$.  Here we chose $\omega_0=\frac{\omega_L+\omega_S}{2}$, with $\omega_L$ ($\omega_S$) is the frequency of  the incoming (scattered) photon. In this
case we can write
\begin{align}
	\label{eq:Enew}
  E_i  &=  \tilde E_i e^{-i\omega_0 t} + \tilde E_i^* e^{i\omega_0 t} ,\\
  \chi_i^\alpha &= \tilde\chi_i^\alpha e^{i\omega_0 t} ,
\end{align}
where $\tilde E_i$ and $\tilde\chi$ are slowly varying fields (e.g.,
fields that vary with frequencies much smaller than $\omega_0$).
Subsituting into the action and dropping the rapidly oscillating
terms, the action can be rewritten as (for notational simplicity we
also drop the tildas in $\tilde E_i$ and $\tilde\chi$)
\begin{multline} \label{eq:Lag00}
  \mathcal{L}=i\psi_\alpha^\dagger \partial_t \psi^\alpha
  -\frac{D_i \psi_\alpha^\dagger D_i\psi^\alpha}{2m^*}
  + e(P^*\psi_\alpha^\dagger \chi^\alpha_i E_i
  +P\chi^\dagger_{i\alpha}\psi^\alpha E_i^*)
  +i\chi^{\dagger}_{i\alpha}\partial_t\chi^\alpha_i 
  +(E_G-\omega_0)\chi^{\dagger}_{i\alpha}\chi^\alpha_i\\
  -
  \chi^{\dagger}_{i\lambda}
  H_{ij}
  \chi^\lambda_j 
  +\lambda^\dagger_\alpha{(\sigma^i)^\alpha}_\beta \chi^\beta_i
  +\chi^\dagger_{i\alpha}{(\sigma^i)^\alpha}_\beta \lambda^\beta .
\end{multline}
The last two terms are Lagrange multiplier for constraints
(\ref{eq:cont}).

Integrating out $\chi$ and $\lambda$ is equivalent to solving the
field equations and the constraints
\begin{align}\label{saddlepoint}
  & eP  \psi^\alpha E_i^*
  + (E_G -\omega) \chi_i^\alpha + {(\sigma^i)^\alpha}_\beta \lambda^\beta
  + i\d_t \chi_i^\alpha - H_{ij} \chi^\alpha_j = 0 ,\\
  & {(\sigma^i)^\alpha}_\beta \chi_i^\beta = 0 .
\end{align}

We will focus on the regime where the photon energy is not too close
to the gap.  More precisely, we will assume that the detuning
$|\omega_0-E_G|$ is still much larger than the distance between the
Landau levels in the bands,
\begin{equation}
  \omega_c \equiv \frac {|e|B}{mc} \ll |\omega_0- E_G| \ll \omega_0 . 
\end{equation}
In the FQH regime that and holes the typical momentum scale is
$1/\ell_B$, one has
\begin{equation}
  H \sim \frac 1m D^2
  \sim \frac{eB}{mc} \,.
\end{equation}
This means that one can solve Eq.~(\ref{saddlepoint}), ignoring the
$\d_t$ and $H$ terms,
\begin{equation}
\label{eq:chi}
  \chi^\alpha_i=-\frac{eP}{3(E_G-\omega_0)}\left[2\psi^\alpha E^*_i+i\epsilon^{ijk}{(\sigma^j)^\alpha}_\beta \psi^\beta E_k^* \right]
   + O\left( \frac{\omega_c}{|E_G-\omega_0|}\right) .
\end{equation}
Substituting this solution into the action, we then find the effective
action for $\psi$ alone.  In fact since $\chi$ is the saddlie point,
an error of order $O(\omega_c/|E_G-\omega_0|)$ translated into an
error $O(\omega_c^2/|E_G-\omega_0|^2)$ in the action; thus it is
justified to also keep terms the terms $\chi^\+\d_t\chi$ and $\chi^\+
H\chi$ when we do the subsitution (\ref{eq:chi}).

To simplify the result, we assume that the electrons are fully
polarized with spin $s_z =\frac12$, so
\begin{equation}
  \psi^\alpha = \begin{pmatrix} \psi \\ 0 \end{pmatrix} .
\end{equation}
We assume the incoming and outgoing photons to have momenta along the
$z$ direction, so $E_i$ are independent of $x$ and $y$.  In this case, the
Lagrangian describing the interaction of the conduction-band electron with the Raman photons have the form
\begin{equation}
  \mathcal I_{\text{int}} = V_{\alpha\beta} (E^*)^\alpha E^\beta ,
\end{equation}
where $V_{\alpha\beta}$ is some operators quadratic over $\psi$.

The photon
spin then points along or opposite to the $z$-axis, corresponding to
the two circular polarizations.  We will distinguish processes in
which the direction of the spin of the photon flips from those in which
the photon spin does not change direction.
Introducing the complex coordinates (in the
quantum Hall convention)
\begin{equation}\label{complex-coord}
  z = x - iy, \quad \bar z = x + iy, \quad
  \d_z = \frac12 (\d_x + i \d_y) , \quad
    \d_{\bar z} = \frac12 (\d_x - i \d_y),
\end{equation}
so
\begin{equation}\label{complex-E}
  E^z = E_x - i E_y, \qquad E^{\bar z} = E_x + i E_y ,
\end{equation}
and
\begin{equation}
  D_z = \frac12 (D_x + iD_y), \qquad D_{\bar z} = \frac12 (D_x -iD_y),
\end{equation}
the interaction Lagrangian can be written as
\begin{equation}
  \mathcal I_{\rm int} = V_{zz} (E^{\bar z})^* E^z
  + V_{\bar z \bar z} (E^z)^* E^{\bar z}
  + V_{z\bar z} (E^{\bar z})^* E^{\bar z}
  + V_{\bar z z} (E^z)^* E^z .
\end{equation}

The terms responsible for scatterings with a switch in the photon helicity
contain $V_{zz}$ and $V_{\bar z\bar z}$.
Direct calculation yields
\begin{align}
  V_{\bar z\bar z} &= \frac{e^2|P|^2}{6(E_G-\omega_0)^2}
  \left[ (\gamma_3+\gamma_2) D_z \psi^\+ D_z\psi
    -  (\gamma_3 - \gamma_2) D_{\bar z} \psi^\+ D_{\bar z}\psi
    \right],\\
  V_{zz} &= \frac{e^2|P|^2}{6(E_G-\omega_0)^2}
  \left[ (\gamma_3+\gamma_2) D_{\bar z} \psi^\+ D_{\bar z}\psi
    -  (\gamma_3 - \gamma_2) D_z \psi^\+ D_z\psi
    \right].
\end{align}
Terms proportional to $\gamma_3+\gamma_2$ in $V_{zz}$ and $V_{\bar
  z\bar z}$ preserves rotational symmetry, while terms proportional to
$\gamma_3-\gamma_2$ breaks the angular momentum conservation by 4.

It is convenient to introduce the ``kinetic'' stress tensor
\begin{equation}
  T_{zz}^{\text{kin}} = \frac 1{m^*} D_z\psi^\+ D_z\psi, \quad
  T_{\bar z\bar z}^{\text{kin}} = \frac1{m^*} D_{\bar z}\psi^\+ D_{\bar z}\psi,
  \quad
  T_{z\bar z}^{\text{kin}} = \frac1{2m^*}
    (D_z\psi^\+ D_{\bar z}\psi + D_{\bar z}\psi^\+ D_z\psi).
\end{equation}
These components are the variation of the kinetic-energy part of the
Hamiltonian over the external metric.  This is different from the full
stress tensor which contains also the variation of the potential
energy over the metric.  The effective operators coupled to the Raman
photon (and flips the direction of its spin) are
\begin{align}
  V_{\bar z \bar z} &= \frac{e^2|P|^2 m^*}{6(E_G-\omega_0)^2}
  \left[ (\gamma_3+\gamma_2) T_{zz}
    -  (\gamma_3 - \gamma_2) T_{\bar z\bar z}
    \right] ,\\
  V_{z z} &= \frac{e^2|P|^2 m^*}{6(E_G-\omega_0)^2}
  \left[ (\gamma_3+\gamma_2) T_{\bar z\bar z}
    -  (\gamma_3 - \gamma_2) T_{zz}
    \right] .
\end{align}
For details of calculations see the Appendix \ref{app:detailed_Raman}.
For completeness, we also write down the operators that do not flip
the photon spin,
\begin{subequations}\label{Vzzbar}
\begin{align}
  V_{\bar{z}z}&=\frac{e^2|P|^2}{(E_G{-}\omega_0)^2}
  \left[\left( \theta^* \omega_c{-}\frac{(E_G{-}\omega_0)}{2}{+}\frac{ \beta^* \omega_c}{4}\right)\rho+\left(2 \alpha^*{-}\beta^*{+}\gamma'^*\right)T_{z \bar{z}}+\frac{i}{2}\psi^\dagger \partial_t \psi\right],\\
  V_{z  \bar z }&=\frac{e^2|P|^2}{9(E_G{-}\omega_0)^2}
  \left[\left( -\theta^* \omega_c{-}\frac{9(E_G{-}\omega_0)}{2}{-}\frac{ \beta^* \omega_c}{4}\right)\rho+\left(6 \alpha^*{-}5\beta^*{+}5\gamma'^*\right)T_{z \bar{z}}+\frac{9i}{2}\psi^\dagger \partial_t \psi\right],
\end{align}
\end{subequations}
where
\begin{align}
\alpha^* &=\frac{m^*}{2m}\left(\gamma_1+\frac{5\gamma_2}{2}\right), \quad
\gamma'^*=\frac{m^*}{m}(\gamma_3-\gamma_2) , \\
\beta^* &= \frac{m^*}{m}\gamma_3,\quad
\theta^*=-\frac{m^*}{m}\left(\kappa+\frac{\gamma_3}{2}\right) .
\end{align}

\section{Scattering of circularly polarized light in the FQH regime}
\label{sec:exp}

\begin{figure}[h!]
 \centering
  \includegraphics[width=0.9 \textwidth]{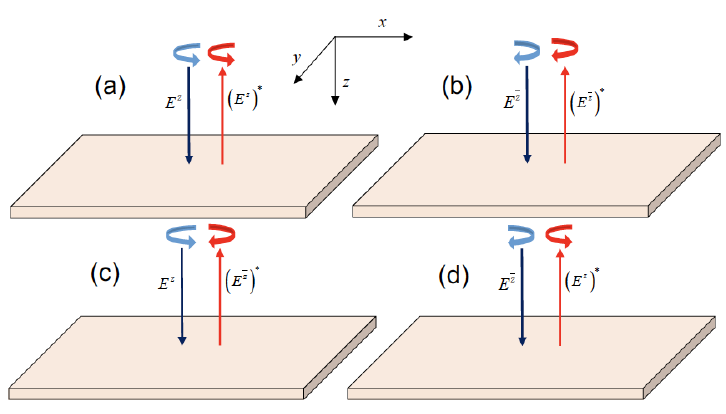}
 \caption{Setup experiment for circular polarized light scattering  }
  \label{fig:Cirl}
 \end{figure}

In this Section, we will calculate the Raman scattering on the fractional quantum Hall state.  The magnitude of the effect can be characterized by the
per-particle differential cross-ssection
\begin{equation}
  \frac{d\sigma_{\lambda\lambda'}}{d\omega d\Omega} \,,
\end{equation}
where $\omega$ is the difference between the energy of the incoming
photon $\omega_L$ and the scattered photon $\omega_S$:
$\omega=\omega_L-\omega_S$, $d\Omega$ is the infinitesimal solid angle
of the scattered photon, and $\lambda$ and $\lambda'$ are the indices
denoting the polarization of the incoming and scattered photons.  For
simplicity, we consider the case when the incident and reflection
light are directed perpendicularly to the sample.  The light can pass
through the sample, or, as depicted in Fig.~\ref{fig:Cirl}, be
reflected from the sample.  
We will assume that both
incident light and scattered lights have circular polarization,
and and $\lambda$ and $\lambda'$ can be either $+$ and $-$ depending on the
projection of the proton spin on the $z$ axis.  For example, for $\sigma_{++}$
the incident light is left-handed (in the ``classical
optics'' convention, see, e.g., Ref.~\cite{Barron2004}) and so the
incident photons have spin pointing along the direction of their
momentum, and the scattered light is right-handed, as in
Fig.~\ref{fig:Cirl} (a)),
we have the formula for cross section per
electron~\cite{PlatzmanHe,Blum},
\begin{equation}
  \frac{d\sigma_{++}(\omega)}{d\omega d\Omega } =
  \frac{1}{N_e}\frac{\omega_S}{\omega_L}\omega_S^2 \omega_L^2
  \sum_f|\langle f |V_{\bar z z}|i\rangle|^2\delta(\epsilon_f-\epsilon_i-\hbar \omega)
  \approx
  -\frac{1}{\bar{\rho}}\frac{\omega_0^4}{\pi}\Im
  \<  V^\+_{\bar zz} V_{\bar zz} \>_{\omega,\mathbf{0}} ,
\end{equation}
with $N_e$ being the total electron number in the conductance band and
$\bar{\rho}$ is the electron density in the conductance band,
$\epsilon_f,\epsilon_i$ are energies of final and initial states,
and
\begin{equation}
  \< A^\+ A \>_{\omega,\k} \equiv \int\!dt\, d\x\, e^{i\omega t-i\k\cdot\x}
  \< T A(t,\x) A^\+(0,\mathbf{0}) \> .
\end{equation}
Thus we need to calculate the spectral density of the operator
$V_{\bar zz}(\omega,0)$.

Similarly, in the case of setups in Figure \ref{fig:Cirl} (b), (c), and
(d), we have
\begin{align}
  \frac{d\sigma_{--}(\omega)}{d\omega d\Omega } &=
  -\frac{1}{\bar{\rho}}\frac{\omega_0^4}{\pi}\Im
  \langle V^\dagger_{z\bar z} V_{z\bar z} \rangle ,
  \\
  \frac{d\sigma_{+-}(\omega)}{d\omega d\Omega } &=
  -\frac{1}{\bar{\rho}}\frac{\omega_0^4}{\pi}\Im
  \langle V^\dagger_{zz} V_{zz} \rangle ,
  \\
  \frac{d\sigma_{-+}(\omega)}{d\omega d\Omega } &=
  -\frac{1}{\bar{\rho}}\frac{\omega_0^4}{\pi}\Im
  \langle V^\dagger_{\bar z\bar{z}} V_{\bar z\bar{z}} \rangle .
\end{align}
The intensity of the Raman scattering in these channels are
proportional to the spectral densities of he operators $V_{z\bar z}$,
$V_{zz}$, and $V_{\bar z\bar z}$.

Let us now show that in the limit of negligible Landau level mixing,
the spectral densities of the operators $V_{z\bar z}$ and $V_{\bar
  zz}$ are zero, implying that the processes depicted on
Fig.~\ref{fig:Cirl} (a) and (b) do not happen.  For that, we note from
Eqs.~(\ref{Vzzbar}) that the integrals of $V_{z\bar z}$ and $V_{\bar
  zz}$ over space are linear combinations of
\begin{equation}
  \int\!d\x\,\rho, \quad \int\!d\x\, T_{z\bar z}^{\text{kin}}, \quad 
  \int\!d\x\, i\psi^\+ \d_t \psi .
\end{equation}
The first integral is the total number of
particles $N_e$.  As this quantity is conserved, it does not contribute to
the spectral density.  From Appendix~\ref{sec:psidtpsi} we find the results  for $N^{th}$ Landau level
\begin{align}
  \int\! d\x\, T_{z\bar z}^{\text{kin}} &= (N+\frac{1}{2})\frac{\omega_c}2 N_e, \\
  \int\! d\x\, i\psi^\+ \d_t \psi &= 2 E-(N+\frac{1}{2})\omega_c N_e,
\end{align}
where $E$ is the total energy.  Both integrals reduce to conserved
quantities.  Thus, the Raman processes that does not involve flipping
the direction of the photon spin are suppressed.

In previous experiments~\cite{Pinczuk1,Pinczuk2}, the momentum
transfer to the electron gass is rather small $kl_B\leq 0.15$.  This
implies that these experiments mainly probe the transitions where the
photon spin flips sign, and effectively measures the spectral
densities of the traceless components of the kinetic stress tensor.
The picture suggested here is different from the previous one
suggested in Ref.~\cite{PlatzmanHe} where the main coupling of the
Raman photon to the electron liquids is through the $\psi^\+\psi
A_i^2$ term in the Lagrangian.  This coupling would lead to a
vanishing Raman scattering at $k=0$.

Let us introduce the short-hand notation for the spectral densities of
the off-diagonal components of the stress tensor,
\begin{align}
\label{eq:Ip}
  -\Im\left\langle T_{zz}^{\text{kin}} T_{\bar{z}\bar{z}}^{\text{kin}} \right\rangle_{\omega,\mathbf{0}} &=I_+(\omega) , \\
\label{eq:Im}
  -\Im\left\langle T_{\bar{z}\bar{z}}^{\text{kin}} T_{zz}^{\text{kin}} \right\rangle_{\omega,\mathbf{0}} &=I_-(\omega) .
\end{align}
These functions should be 
calculated numerically.  The intensity of Raman scatterings can now be
expressed as
\begin{align}
  \frac{d\sigma_{+-}}{d\omega d\Omega}= &
  \frac{1}{\pi\bar \rho}
  \left[ \frac{e^2|P|^2 \omega_0^2 m^*}{6(E_{0}-\omega_{0})^2} \right]^2
  \left[(\gamma_3+\gamma_2)^2I_+(\omega) +(\gamma_3-\gamma_2)^{2}I_-(\omega) \right],\\
    \frac{d\sigma_{-+}}{d\omega d\Omega}= & \frac{1}{\pi\bar \rho}
    \left[ \frac{e^2|P|^2 \omega_0^2 m^*}{6(E_{0}-\omega_{0})^2} \right]^2\left[(\gamma_3+\gamma_2)^2I_-(\omega) +(\gamma_3-\gamma_2)^{2}I_+(\omega)
      \right] . \end{align}

In Ref.~\cite{Nguyen2014LLL} it was proven that
$I_+(\omega)=0$ for the trial ground states of model Hamiltonians with
contact interactions.  While there is no argument that $I_+$ should be zero
for more general Hamiltonians, 
numerically it was found that for Coulomb interaction $I_+$ is much smaller than
$I_-$ for the Laughlin $\nu=1/3$ state 
\cite{Liou2019}. 

If one ignore $I_+$ compared to $I_-$, 
we find the ratio of scattered light intensity of experiment setups \ref{fig:Cirl} (c) and  \ref{fig:Cirl} (d)
\begin{equation}
\label{eq:ratio}
\frac{I_{-+}(\omega=\Delta)}{I_{+-}(\omega=\Delta)}=\frac{(\gamma_3+\gamma_2)^{2}}{(\gamma_3-\gamma_2)^{2}} \,.
\end{equation}
The ratio only depends on the Luttinger parameters. Moreover, the fact that $I_{+-}$ will vanish if $\gamma_3-\gamma_2=0$ suggests that the signal of $I_{+-}$ is due to rotational symmetry breaking. These results confirm that at zero momentum ($\mathbf{k}= 0$), the magneto-roton excitation has spin 2 in $\hat{z}$ direction. However, in the case of finite momentum, the magneto-roton excitation will be mixed of modes with spin $+2$ and spin $-2$ in $\hat{z}$ direction, which was suggested in the previous work \cite{Golkar}. \\

The numerical values for the Luttinger parameters of GaAs are~\cite{Ruf} 
\begin{equation}
\gamma_1=6.9,\quad
\gamma_2=2.1,\quad
\gamma_3=2.9,\quad
\kappa=1.2 .
\end{equation}
Substituting these parameters in equation (\ref{eq:ratio}) yields the ratio of intensities
\begin{equation}
\frac{I_{-+}(\omega=\Delta)}{I_{+-}(\omega=\Delta)}\approx 40 .
\end{equation}
Note that this relies on the assumption that $I_+=0$, which is not
expected to hold exactly for the Coulomb interaction.  However, if
$I_+$ is small compared to $I_-$, one still expect that $I_{-+}\ll
I_{+-}$ for $\nu=1/3$ states.
This is also expected for the Jain states $\nu=n/(2n+1)$, in which the
composite fermion theory implies that the magnetoroton has the same sign of
spin as in the $\nu=1/3$ state.  In the particle-hole conjugate Jain states
$\nu=(n+1)/(2n+1)$, in contrast, one expects that
$I_{-+}\gg I_{+-}$~\cite{Golkar:2016thq,Nguyen:2017mcw}.


\section{Stress tensor projected on a Landau level }
\label{sec:TLL}

As we have seen from the previous section, to obtain the cross section
of polarized Raman scattering, we need to calculate the spectral
function of the kinetic part of the stress tensor projected on a
specific Landau level (the fractionally filled one).  To enable future
numerical calculations of these spectral functions we need the
expressions for the operators $T^{\text{kin}}_{ij}$ after the
projection to a Landau level.  In this section, we will derive the
explicit form of the projected kinetic stress tensor.

We summarize the result here.  For a system of particles interacting
through a two-body isotropic potential $V(|\x-\y|)$ on the $N$th
Landau level, the kinetic part of the stress tensor (at zero momentum)
can be written as
\begin{equation}\label{Tkin-summarized}
  \int\! d\x\, T_{ij}^{\text{kin}} =
  \int_\q \frac{q_i q_j}{q} \frac\d{\d q} \left\{ e^{-x_q} [L_N(x_q)]^2
  \right\}
  V(q) \bar \rho(\q) \bar \rho(-\q),
\end{equation}
where $\int_\q\equiv \int\!d\q/(2\pi)^2$,
\begin{equation}
  x_q \equiv \frac{q^2\ell_B^2}2\,,
\end{equation}
$L_N(x)$ is the Laguerre polynomial, and $ij$ can be either $zz$ or
$\bar z\bar z$, and $\bar\rho(\q)$ is the projected density operator
in momentum space~\cite{Girvin1986}.

The interpretation of the above equation is rather simple.  Recall
that the projected Hamiltonian of the system is
\begin{equation}
  H = \int_\q  e^{-x_q} [L_N(x_q)]^2
  V(q) \bar \rho(\q) \bar \rho(-\q),
\end{equation}
where the form-factor $e^{-x_q}[L_N(x_q)]^2$ arises from the
projection to the $N$th Landau level.  Polarized Raman scattering, as
explained above, has the effect of changing the effective metric in
the kinetic term for the electron (making the effective mass $m^*$
anisotropic).  This makes the Landau orbit on the $N$th Landau level
anistropic, and the effect of that is the operator $(q_iq_j/q)\d_q$
acting on the form-factor.

For $N=0$, Eq.~(\ref{Tkin-summarized}) reads
\begin{equation}
  \int\! d\x\, T_{zz}^{\text{kin}} = -\ell_B^2
  \int_\q q_z^2 e^{-q^2\ell_B^2/2}
  V(q) \bar \rho(\q) \bar \rho(-\q) ,
\end{equation}
which is exactly the operator considered in Ref.~\cite{Liou2019}.
Thus the spectral densities computed in Ref.~\cite{Liou2019} are
directly related to polarized Raman scattering on FQH states on the
LLL.

For the next-to-lowest Landau level $N=1$ we have
\begin{equation}
  \int\! d\x\, T_{zz}^{\text{kin}} = -\ell_B^2
  \int_\q q_z^2 e^{-x_q} (1-x_q)(3-x_q)
  V(q) \bar \rho(\q) \bar \rho(-\q).
\end{equation}
The general expression for the kinetic stress
tensor~(\ref{Tkin-summarized}) has also been found by Kun
Yang~\cite{KunYang-unpublished}.
In the rest of this Section, we provide a derivation of
Eq.~(\ref{Tkin-summarized}).

\subsection{Preliminaries}

We use the complex coordinates~(\ref{complex-coord}) 
and the symmetric gauge $A_x=-\frac12By$, $A_y=\frac12Bx$.  In the complex
coordinates
\begin{equation}
  A_z = \frac12 (A_x + i A_y) = i \frac B4 \bar z,\qquad
  A_{\bar z} = \frac12 (A_x - i A_y) = -i \frac B4 z,
\end{equation}
Then in the symmetric gauge
\begin{align}
  D_z &= \d_z - \frac{ie}c A_z = \d_z - \frac{\bar z}4 \,,\\
  D_{\bar z} & = \d_{\bar z} - \frac{ie}c A_{\bar z} = \d_{\bar z} + \frac z4 
  \,.
\end{align}
Note that
\begin{equation}
	\label{eq:comm}
  [D_z, \, D_{\bar z}] = -\frac{eB}{2c} = \frac1{2\ell_B^2} \,.
\end{equation}
The (complex) guiding center coordinates are defined as
\begin{align}
  Z &= z - 2 \ell_B^2 D_{\bar z} = \frac z2 - 2 \ell_B^2\d_{\bar z}, \\
  \bar Z &= \bar z + 2 \ell_B^2 D_z = \frac{\bar z}2 + 2 \ell_B^2\d_z
\end{align}
which satisfy
\begin{equation}\label{DzZcomm}
  [D_z, \, Z] = [D_z, \, \bar Z]
  = [D_{\bar z}, \, Z] = [D_{\bar z}, \, \bar Z] = 0 .
\end{equation}
and
\begin{equation}
  [\bar Z, \, Z] = 2 \ell_B^2 .
\end{equation}
We define another set of coordinates: the relative coordinates which
describes the motion around the guiding center,
\begin{align}
  \zeta &= 2 \ell_B^2 D_{\bar z} = \frac z2 + 2\ell_B^2\d_{\bar z} ,\\
  \bar \zeta &= -2\ell_B^2 D_z = \frac{\bar z}2 -2\ell_B^2\d_z ,
\end{align}
which commute with $Z$ and $\bar Z$ [Eqs.~(\ref{DzZcomm})] and have
the commutator
\begin{equation}
  [\bar\zeta,\, \zeta] = -2\ell_B^2.
\end{equation}
Then $z=Z+\zeta$, $\bar z = \bar Z+\bar\zeta$.  We denote the 2D
vector whose complex coordinates are $Z$ and $\bar Z$ as $\R$, and the
vector with complex coordinates $\zeta$ and $\bar\zeta$ as
$\tilde{\mathbf{r}}$.  That means $\x=\R+\tilde{\mathbf{r}}$.

One defines two sets of creation and annihilation operators.  One set
moves between different Landau levels
\begin{equation}
  a = \sqrt2 \ell_B D_{\bar z} = \frac\zeta{\sqrt2 \ell_B} , \qquad
  a^\+ = - \sqrt2 \ell_B D_z = \frac{\bar\zeta}{\sqrt2 \ell_B} \,,
\end{equation}
and another set moves within a Landau level
\begin{equation}
  b = \frac1{\sqrt2 \ell_B} \bar Z, \qquad b^\+ = \frac1{\sqrt2 \ell_B} Z .
\end{equation}
The orbitals are obtained by acting creation operator on the lowest state
\begin{equation}
  | M,m\> = \frac1{\sqrt{M!m!}} a^{\+M} b^{\+m} |0,0\>,
\end{equation}
where 
\begin{equation}
  \<\x|0,0\> \sim e^{-|z|^2/4\ell_B^2}.
\end{equation}

\subsection{The kinetic stress tensor on a Landau level}

Our task is to find the expression for the kinetic part of the stress
tensor in the theory where the electrons live on one Landau level.
This will be done through a field-theory formalism.  The action
describing electrons on the $N$th Landau level is
\begin{multline}
  S = \int\!dt\, d\x \left[ i \psi^\+ \d_t \psi
      + \chi^\+ \left( 2 \ell_B^2D_z D_{\bar z}\psi + N \psi \right) 
      + \left( 2 \ell_B^2D_{\bar z} D_z\psi^\+ + N \psi^\+ \right)\chi \right]\\
      - \frac12 \!\int\!dt\, d\x\, d\x'\, V(\x-\x') \psi^\+(\x) \psi^\+(\x')
      \psi(\x')\psi(\x) .
\end{multline}
The fields $\chi$ and $\chi^\+$ are simply the Lagrange multipliers enforcing
the constraint
\begin{equation}
  2\ell_B^2  D_z D_{\bar z} \psi + N\psi  = 0,  
\end{equation}
which is simply the condition that $\psi$ lies on the $N$th Landau
level.

To find the stress tensor, we first rewrite the action by integration
by part,
\begin{multline}
  S = \int\!dt\, d\x \left[ i \psi^\+ \d_t \psi
      -2 \ell_B^2D_z \chi^\+  D_{\bar z}\psi  
      -  2 \ell_B^2D_z\psi^\+ D_{\bar z}\chi
      + N (\chi^\+\psi+\psi^\+ \chi) \right]\\
      - \frac12 \!\int\!dt\, d\x\, d\x'\, V(\x-\x') \psi^\+(\x) \psi^\+(\x')
      \psi(\x')\psi(\x) ,
\end{multline}
then the kinetic part of the stress tensor can be calculated from
Noether's theorem:
\begin{equation}
  {T^i}_j = - \frac{\d L}{\d (\d_i\phi_a)} \d_j\phi_a ,
\end{equation}
where one sums over all fields $\phi_a$, which in our case encompass
$\psi$, $\phi^\+$, $\chi$, and $\chi^\+$.  For the polarized Raman
experiment with perpendicularly incoming and outgoing photons, with a
flipping of the photon spin, one only needs the traceless part of the
stress tensor, integrated over space:
\begin{align}\label{intT}
  \int\!d\x\, T^{\text{kin}}_{zz} &= -\ell_B^2\int\!d\x\, (\chi^\+ D_z^2\psi + D_z^2\psi^\+ \chi),\\
  \int\!d\x\, T^{\text{kin}}_{\bar z\bar z} &= -\ell_B^2\int\!d\x\, (\chi^\+ D_{\bar z}^2\psi
  + D_{\bar z}^2\psi^\+ \chi).
\end{align}
We can expand $\chi$ as a sum over Landau levels:
$\chi=\chi_0+\chi_1+\chi_2+\cdots$.  We recall that when acting on
$\psi$ and $\chi$, $D_z$ raises and $D_{\bar z}$ lowers the Landau
level index, while when acting on $\psi^\+$ and $\chi^\+$ they switch
roles.  Due to the orthogonality of wavefunctions on different Landau
levels, only the parts of $\chi$ that are on the $(N+2)$th and (if
$N\ge2$) $(N-2)$th Landau level's contribute to the integrals in
Eqs.~(\ref{intT}).
We then have, for $N\ge2$ 
\begin{align}
  \int\!d\x\, T^{\text{kin}}_{zz} &= -\ell_B^2\int\!d\x\, (\chi_{N+2}^\+ D_z^2\psi + D_z^2\psi^\+ \chi_{N-2}) ,\\
  \int\!d\x\, T^{\text{kin}}_{\bar z\bar z} &= -\ell_B^2\int\!d\x\, (\chi_{N-2}^\+ D_{\bar z}^2\psi
  + D_{\bar z}^2\psi^\+ \chi_{N+2}) ,
\end{align}
and for $N=0$ or 1
\begin{align}
  \int\!d\x\, T^{\text{kin}}_{zz} &= -\ell_B^2\int\!d\x\, \chi_{N+2}^\+ D_z^2\psi , \\
  \int\!d\x\, T^{\text{kin}}_{\bar z\bar z} &= -\ell_B^2\int\!d\x\,
  D_{\bar z}^2\psi^\+ \chi_{N+2} .
\end{align}
The equation determining $\chi$ is
\begin{equation}\label{chi-eq1}
  0 = \frac{\delta S}{\delta \psi^\+} =
  i \d_t \psi + ( 2 \ell_B^2 D_z D_{\bar z} + N) \chi +  W(\x), 
\end{equation}
which, for $n\neq N$, implies
\begin{align}
  \chi_n = \frac{W_n}{n-N} \,.
\end{align}
In particular
\begin{align}
  \chi_{N+2} &= \frac12 W_{N+2}\,,\\
  \chi_{N-2} &= -\frac12 W_{N-2}\, \quad (N\ge2),
\end{align}
and therefore 
\begin{align}
  \int\!d\x\, T^{\text{kin}}_{zz} &= -\frac12 \ell_B^2\int\!d\x\, (W^\+ D_z^2\psi - D_z^2\psi^\+ W) , \\
  \int\!d\x\, T^{\text{kin}}_{\bar z\bar z} &= \frac12 \ell_B^2\int\!d\x\, (W^\+ D_{\bar z}^2\psi
  - D_{\bar z}^2\psi^\+ W) ,
\end{align}
where we have used the orthogonality of the functions on different
Landau levels to replace $W_{N+2}$ and $W_{N-2}$ by simply $W$.

Using formulas of Appendix~\ref{sec:project}, we then find
\begin{equation}
  \int\!d\x\, T^{\text{kin}}_{zz} =
  - \frac {1}{2} \int_{\q} \ell_B^2q_z^2 e^{-x_q}
  L_N ( x_q)
  \left[ L_N^2 ( x_q ) - L_{N-2}^2 ( x_q ) \right]
  V(q) \bar \rho(\q) \bar \rho(-\q),  \quad x_q \equiv \frac{q^2\ell_B^2}2\,,
\end{equation}
and a similar equation where $T_{zz}$ is replaced by $T_{\bar z\bar
  z}$ and $q_z$ by $q_{\bar z}$.  Here $L_N$ is the Laguerre
polynomial and $L_N^2$ is not the square of $L_N$ but the associate
Laguerre polynomial $L_N^k$ with $k=2$, and for the uniformity of the
equation we have defined $L^2_{-1}=L^2_{-2}=0$.

These equations can be brought to an alternative form by using the
following identities involving the associated Laguerre polynomials,
\begin{equation}
  L_N^k(x)=L_N^{k+1}(x)-L_{N-1}^{k+1}(x), \qquad
  \frac{d}{dx}L^{k}_N(x)=-L_{N-1}^{k+1}(x) .
\end{equation}
One can show that, for $N\geq 2$ 
\begin{equation}
	L_N^2(x) - L_{N-2}^2(x)=L_N(x)-2 \frac{d}{dx}L_N(x),
\end{equation}
while one can also check directly that
\begin{equation}
	L_0(x)\left[L_0(x)-2L'_0(x)\right]=L_0(x)L_0^2(x), \qquad L_1(x)\left[L_1(x)-2L'_1(x)\right]= L_1(x)L_1^2(x).
\end{equation}
We then can rewrite the kinetic part of stress tensor for a general
Landau level $N$ as
\begin{equation}\label{eq:Tzkin}
  \int\!d\x\, T^{\rm kin}_{zz} = - \frac {\ell_B^2}{2 } 
	\sum_{\q}
	q_z^2 e^{-x_q }L_N(x_q)\left[L_N(x_q)-2L'_N(x_q)\right]V(q)  \bar\rho(\q) \bar\rho(-\q),
\end{equation}
and another equation with the replacement $T_{zz}^{\text{kin}}\to
T_{\bar z\bar z}^{\text{kin}}$ and $q_z\to q_{\bar z}$.  This can
be further transformed to Eq.~(\ref{Tkin-summarized}).

Some remarks are in order.  The kinetic part of stress tensor
operators \eqref{eq:Tzkin} for the LLL share the same form as the
spin-2 operators in Ref~\cite{Liou2019}, in which the authors
calculated the normalized spectral functions. One can employ the same
approach to obtain the spectral density of the stress tensor for
higher Landau levels. The result will provide the estimation for Raman
scattering intensity of an FQH system at higher Landau levels in our
theoretical model.  In Appendix \ref{sec:sumrule}, we give the
expression for the full stress tensor operators, including the
contribution from the interaction.  This can be used calculate the
spectral function of LLL stress tensor and check the sum rules derived
in Ref.~\cite{Golkar}.


\section{Conclusions}
\label{sec:concl}

In this paper, we have derived the coupling of the electrons in a single Landau
level with applied electromagnetic waves, which effectively
captures the essential physics of Raman scattering on FQH
systems.  We show that the electron operator responsible for Raman scattering
is not the density operator, but the ``kinetic stress tensor,'' and we 
derive the expression of the latter after projection to a single
Landau level. We then show that, in the long-wavelength regime, the
light scattering intensity in Raman experiments measures the spectral
function of the kinetic part of stress tensors.  Our
calculation explains the scattering intensity peaks at
zero momentum without relying on any momentum-noncoserving processes.

In addition, we proposed experimental setups to verify the spin-2
hypothesis of magneto-roton mode in FQH systems using Raman scattering
with circularly polarized light.  We show that, for a magnetoroton
with a well-defined sign of spin, the ratio between light scattering
intensities of different configurations of circularly polarized Raman
experiments only depends on Luttinger parameters, which are well
known.  Measuring those ratios can confirm our theoretical model and
unveil the spin of the magnetoroton excitations in a FQH state.

Using the explicit form of the stress tensor operator derived in this
paper, one can perform the numerical calculation to obtain the stress
tensors' spectral function.  One then use the numerical results to
verify the LLL sum rules proposed in Ref.~\cite{Golkar}, and to predict
the result of Raman scattering on states on higher Landau levels.

Raman scattering may help resolve the question about the nature of the
$\nu=5/2$ state.  In a recent
experiment~\cite{BanerjeeThermalHall:2018}, the thermal Hall
conductance at the edge of the $\nu=5/2$ state was determined to be
consistent with the PH-Pfaffian state~\cite{Son:2015xqa}, but not the
Pfaffian~\cite{MOOREREAD}, or the anti-Pfaffian
state~\cite{Levin-antiPf,Sungsik-antiPf}, seemingly contradicting the
results of numerical simulations~\cite{Rezayi:2017uac}.  Theoretical
proposals aiming to explain this discrepancy include a
disorder-stabilized thermal metal phase which is adiabatically
connected to the PH-Pfaffian phase~\cite{Mross:2018meb,Wang:2017ecp}
and an incomplete thermalization on the
edge~\cite{SimonThermalHall:2018,MaFeldmanThermalHall:2019,SimonThermalHall:2020}.
Raman scattering provides a way to probe directly the bulk of the
$\nu=5/2$ state.  The magnetoroton in the Pfaffian (Moore-Read)
state~\cite{MOOREREAD} must have a spin of the same sign as in the
$\nu=1/3$ Lauglin state, while in the anti-Pfaffian
state~\cite{Levin-antiPf,Sungsik-antiPf} it must have the opposite
sign.  The PH-Pfaffian state~\cite{Son:2015xqa}, in the absence of
Landau-level mixing, is particle-hole symmetric, hence the Raman
scattering probabilities $I_{+-}$ and $I_{-+}$ must be the same. However, it is not clear how significant the effect of Landau level mixing would be in this case.

To derive the coupling of the Raman photons to FQH electron liquid, we
have assume that the detuning $|\omega_0-E_G|$ is much larger than the
cyclotron energy $\omega_c$.  This allows us to perform the first step
of our ``factorization'' procedure---integrating out the
holes---without having to think about the effect of the magnetic field
on the conduction-band electrons.  We suspect that our final result is
valid under a weaker assumption---that the detuning is larger than the
energy scale of the FQHE, i.e., of the Coulomb interaction between the
conduction-band electrons.  A derivation of this result would need to
be a one-step procedure---integrating out the valence bands and the
projecting to one Landau level at the same time.  We defer this to
future work.

\acknowledgments

The authors thank Duncan Haldane, Edward Rezayi, and Kun Yang for
discussion.  This work is supported, in part, by by the U.S.\ DOE
grant No.\ DE-FG02-13ER41958, a Simons Investigator grant and by the
Simons Collaboration on Ultra-Quantum Matter, which is a grant from
the Simons Foundation (651440, DTS). DXN
was supported by Brown Theoretical Physics Center.  

\appendix

\section{Detailed derivation of Raman scattering coupling}
\label{app:detailed_Raman}

In this section, we present the detailed derivation the coupling of the FQH system with photon. We define new parameters
\begin{align}
  \alpha &= \frac{1}{2m}\left(\gamma_1+\frac{5\gamma_2}{2}\right), \quad
  \gamma' = \frac{1}{m}(\gamma_3-\gamma_2),\\
  \beta &= \frac{1}{m}\gamma_3 ,\quad
  \theta = -\frac{1}{m}\left(\kappa+\frac{\gamma_3}{2}\right).
\end{align}

After integrating out fields in valence band, we derive the effective Lagrangian for conduction band
\begin{multline}\label{eq:ef1}
\mathcal{L}=i\psi_\alpha^\dagger \partial_t \psi^\alpha-\frac{D_i\psi_\alpha^\dagger D_i\psi^\alpha}{2m^*}+i\chi^{\dagger}_{i\alpha}\partial_t\chi^\alpha_i -(E_G-\omega_0)\chi^{\dagger}_{i\alpha}\chi^\alpha_i\\
  - \chi^{\dagger}_{k'\lambda}
  \biggl[ \alpha \mathbf{D}^2 +\gamma' \sum_i J_i^2 D_i^2
    -\beta(\J\cdot\mathbf{D})^2  -\frac{e}{c}\theta\J\cdot\B\biggr]_{k'k}
  \chi^\lambda_k .
\end{multline}
 All terms which contain valence band field $\chi^\alpha_i$ can be considered as coupling of conduction band field $\psi^\alpha$ with the electric field through substitution (\ref{eq:chi}). We define
\begin{align}
   \mathcal{I}_0 &=i\chi_{i \lambda}^\dagger \partial_t \chi_i^\lambda , \label{eq:I0}\\
   \mathcal{I}_1 &=-(E_G-\omega_0)\chi^{\dagger}_{i\alpha}\chi^\alpha_i , \label{eq:I1} \\
   \mathcal{I}_\alpha &= -\alpha\chi^{\dagger}_{k\lambda} \mathbf{D}^2\chi^\lambda_k , \\
   \mathcal{I}_\beta &= \beta \chi^{\dagger}_{k'\lambda}
   \left[(\J\cdot\mathbf{D})^2 \right]_{k'k} \chi^\lambda_k , \\ 
  \mathcal{I}_{\gamma'} &= -\gamma' \chi^{\dagger}_{k'\lambda}
  \sum_i (J_i^2)_{k'k} D_i^2 \chi^\lambda_k , \\
  \mathcal{I}_{\theta} &= \theta \frac{e}{c}\sum_{k,k',\lambda} \chi^{\dagger}_{k'\lambda} (\J_{k'k}\cdot \B) \chi^\lambda_k .
\end{align}
Consequently, the effective Lagrangian can be rewritten as
\begin{equation}\label{eq:LinL}
  \mathcal{L}_{\rm eff}=i\psi^\dagger \partial_t \psi
  -\frac{(D_i\psi)^\dagger D_i\psi}{2m^*}+\mathcal{I}_0+\mathcal{I}_1
  +\mathcal{I}_\alpha+\mathcal{I}_\beta+\mathcal{I}_{\gamma'}
  +\mathcal{I}_{\theta} .
\end{equation} 
 Substitution of equation (\ref{eq:chi}) for $\chi^\alpha_i$ in equation (\ref{eq:I1})  yields
\begin{equation}
\mathcal{I}_1=-\frac{e^2|P|^2}{3(E_G-\omega_0)}\left[2 \psi^\dagger \psi |E|^2+i\epsilon^{ijk}\psi^\dagger \sigma^j \psi E_k^*E_i \right] .
\end{equation} 
The first term in $\mathcal{I}_1$ is the interaction of light with
charge density, the second term is the interaction of light with spin
density. Considering that the electrons in the conduction band, under
strong magnetic field in $\hat{z}$ direction, only have the spin
component $s_z=\frac12$, we can rewrite
\begin{equation}
\mathcal{I}_1=-\frac{e^2|P|^2}{3(E_G-\omega_0)}\left[2 E_i E_i^*+i(E_2^*E_1-E_1^*E_2) \right]\rho,
\end{equation}
where $\rho=\psi^\dagger \psi$.   Since $E_i^*$ is a slow varying field under the redefinition \eqref{eq:Enew}, the term with $\partial_t E_i^*$ in $\mathcal{I}_0$ is small in comparison with $\mathcal{I}_1$. We then have
\begin{equation}
	\mathcal{I}_0=i\frac{e^2|P|^2}{3(E_G-\omega_0)^2}\left[2 E_i E_i^*+i(E_2^*E_1-E_1^*E_2) \right]\psi^\dagger \partial_t \psi.
\end{equation}

 To understand the next interaction terms (the Luttinger terms), we recall the formula for the kinetic part of stress energy tensor 
\begin{equation}
  T^{\rm kin}_{ij}=\frac{(D_i \psi)^\dagger D_j \psi}{2m^*}
  +\frac{(D_j \psi)^\dagger D_i \psi}{2m^*}.
\end{equation}
Under above assumption of spin state of electrons in the conduction band and $k_3=0$ (we consider 2D system in $xy$ plane, and the applied magnetic field is in $\hat{z}$ direction), we can rewrite the Luttinger terms as 
\begin{equation}
	\mathcal{I}_\alpha=\frac{\alpha^* e^2|P|^2}{3(E_G-\omega_0)^2}\left[2 E_i E_i^* + i (E_2^*E_1-E_1^*E_2) \right] (T^{\rm kin}_{11}+T^{\rm kin}_{22}) ,
\end{equation}
\begin{equation}
	\begin{aligned}
		\mathcal{I}_\beta=&-\frac{\beta ^* e^2|P|^2}{9(E_G-\omega_0)^2}\left\{T^{\rm kin}_{11}\left[5 E_2 E_2^*+5 E_3 E_3^*+2E_1 E_1^*+i (E_2 E_1^*-E_1E_2^*)  \right]\right. \\
		&\left. +T^{\rm kin}_{22}\left[5 E_1 E_1^*+5 E_3 E_3^*+2E_2 E_2^*+i (E_2 E_1^*-E_1E_2^*)  \right] \right. \\
		&\left. -3 T^{\rm kin}_{12}\left(E_1 E_2^* +E_2 E_1^*\right)+i\frac{\omega_c}{2 c}\left[5(E_2 E_1^*-E_1 E_2^*)-4i E_a E_a^*-2i E_3 E_3^*\right]\rho \right\} ,
	\end{aligned}
\end{equation}
\begin{align}
	\mathcal{I}_{\gamma'}=&\frac{{\gamma' }^* e^2 |P|^2}{9(E_G-\omega_0)^2}\left\{T^{\rm kin}_{11}\left[5 E_2 E_2^*+5 E_3 E_3^*+2E_1 E_1^*+i (E_2 E_1^*-E_1E_2^*)  \right]\right. \nonumber\\
	&\left. +T^{\rm kin}_{22}\left[5 E_1 E_1^*+5 E_3 E_3^*+2E_2 E_2^*+i (E_2 E_1^*-E_1E_2^*)  \right]  
	\right\} ,
\end{align}
\begin{equation}
	\mathcal{I}_{\theta}=i\theta^* \omega_c\frac{ e^2 |P|^2}{9(E_G-\omega_0)^2}\left[5(E_1E_2^*-E_2 E_1^*)+4i E_a E_a^* +2i E_3 E_3^*\right]\rho,
\end{equation}
where we have defined the parameters
\begin{eqnarray}
\beta^*=\beta m^*, \quad
\alpha^*=\alpha m^*,\\
\gamma^{\prime *}=\gamma' m^*, \quad
\theta^*=\theta m^*,
\end{eqnarray}
and the cyclotron frequency 
\begin{equation}
\omega_c=-\frac{eB}{c m^*}=\frac{|eB|}{cm^*}.
\end{equation}
The effective Lagrangian for the conduction band includes the coupling of the electric field $E_i$ with charge density $\rho$ and the kinetic part of stress energy tensor $T^{\rm kin}_{ij}$.
We have the effective interaction of conduction band with light through $\mathcal{I}_0,\mathcal{I}_1,\mathcal{I}_\alpha,\mathcal{I}_\beta, \mathcal{I}_{\gamma'},\mathcal{I}_\theta$. We can rewrite the interaction term in the convenient form for circular polarized light scattering experiment setup in the Figure \ref{fig:Cirl}. In this case, we can consider $E_3=0,E_3^*=0$.
Going to the complex coordinates~(\ref{complex-coord}) and (\ref{complex-E}),
in which
\begin{equation}
	T^{\rm kin}_{zz}=\frac{1}{4}(T^{\rm kin}_{xx}-T^{\rm kin}_{yy}+2iT^{\rm kin}_{xy}),\quad
	T^{\rm kin}_{\bar{z}\bar{z}}=\frac{1}{4}(T^{\rm kin}_{xx}-T^{\rm kin}_{yy}-2iT^{\rm kin}_{xy}),\quad
	T^{\rm kin}_{z\bar{z}}=\frac{1}{4}(T^{\rm kin}_{xx}+T^{\rm kin}_{yy}),
\end{equation}
we can rewrite the interaction terms as 
\begin{equation}
	\label{eq:I0psi2}
	\mathcal{I}_0=i\frac{e^2|P|^2}{6(E_G-\omega_0)^2}\left[3 E^{\bar z} (E^{\bar z })^*+E^z (E^{z})^*\right] \psi^\dagger \partial_t \psi,
\end{equation}
\begin{equation}
	\label{eq:I1psi2}
	\mathcal{I}_1=-\frac{e^2|P|^2}{6(E_G-\omega_0)}\left[3 E^{\bar z}  (E^{\bar z })^*+E^z (E^{z})^*\right] \rho,
\end{equation}
\begin{equation}
	\label{eq:Ibetasi2}
	\mathcal{I}_\alpha=\alpha^*\frac{e^2|P|^2}{3(E_G-\omega_0)^2}\left[6 E^{\bar z} (E^{\bar z })^*+2E^z(E^{z})^*\right] T^{\rm kin}_{z \bar{z}},
\end{equation}
\begin{multline}
	\label{eq:Ibetapsi2}
	\mathcal{I}_\beta=-\frac{\beta^*  e^2|P|^2}{9(E_G-\omega_0)^2}\left\{ -3E^{z}  (E^{\bar z })^* T^{\rm kin}_{zz}-3 E^{\bar{z}}(E^{z})^* T^{\rm kin}_{\bar{z}\bar{z}}+(5 E^{ z}(E^{z})^*+9 E^{\bar z} (E^{\bar z })^* )T^{\rm kin}_{z \bar{z}} \right. \\ 
\left.	+\frac{ \omega_c}{4}(E^{z}(E^{z})^*-9E^{\bar{z}} (E^{\bar z })^*)\rho\right\} ,
\end{multline}
\begin{align}
	\label{eq:Igammapsi2}
	\mathcal{I}_{\gamma'}
	=&\frac{\gamma'^* e^2|P|^2}{9(E_G-\omega_0)^2}
        \left\{-\frac{3}{2}\left[E^z (E^{\bar{z}})^*+E^{\bar z}(E^z)^*\right](T^{\rm kin}_{zz}+T^{\rm kin}_{\bar{z}\bar{z}})+\left[5 E^{z} (E^{z})^*+9 E^{\bar{z}} (E^{\bar z })^* \right]T^{\rm kin}_{z \bar{z}} 
	\right\} ,
\end{align}
\begin{equation}
	\label{eq:Ithetapsi2}
	\mathcal{I}_{\theta}=\theta^* \frac{\omega_c e^2|P|^2}{18(E_G-\omega_0)^2}\left[E^{z}(E^{z})^*-9E^{\bar z} (E^{\bar z})^*  \right]\rho.
\end{equation}
We can easily check that only $\mathcal{I}_{\gamma'}$ violates rotational invariance.

\section{The dipole-transition oefficient $P$}

In this Appendix, we follow Ref.~\cite{Haus2016} to derive an
expression for the dipole-transition coefficient $P$ through the
electron Bloch wave functions. The first term of Eq.~\eqref{eq:photon}
absorbs a photon and creates a hole in the valence band and adds an
electron to the conductance band. We can rewrite this term in the
Hamiltonian as
\begin{equation}
	\label{eq:photon1}
	-\int\! d x \, e P^*_{ij}\psi^{\alpha\dagger} \chi^{i}_{\alpha}E^j,
\end{equation}
with $P_{ij}^*=P^* \delta_{ij}$. Comparing with Eq. (11.23) in Ref.~\cite{Haus2016}, we see that \footnote{We use the Coulomb gauge $A_0=0$ in this section.}
\begin{equation}
-eP_{ij}^* i\omega_{ph}\left< n_{ph}{-}1| A^j |n_{ph} \right>=-\frac{e}{m}\left< \psi^\alpha_\k, n_{ph}{-}1| A^j p_j|\chi_{i\alpha, \k}, n_{ph} \right>,
\end{equation}
with $m$ being the free electron mass and $p_j$ is the free electron momentum operator. Then we have 
\begin{equation}
	P^*_{ij}=-\frac{i}{m \omega_{ph}}\left< \psi^\alpha_\k|  p_j|\chi_{i\alpha,\k} \right>,
\end{equation}
where $\psi^\alpha_\k$ is the Bloch wavefunction of electron in the conductance band ($s$-band) 
\begin{equation}
\psi^\alpha_\k=\frac{1}{\sqrt{N_{\rm site}}}\sum_a e^{i \k \cdot \x} u^\alpha_c(\x-\mathbf{R}_a),
\end{equation}
and $\chi_{i\alpha,\k}$ is the Bloch wavefunction of electron in the valence bands (p-band)
\begin{equation}
	\chi_{i\alpha,\k}=\frac{1}{\sqrt{N_{\rm site}}}\sum_a e^{i \k \cdot \x} u^{i \alpha}_{v}(\x-\mathbf{R}_a).
\end{equation}
We have 
\begin{equation}
  \left< \psi^\alpha_\k|  p_j|\chi_{i\alpha,\k} \right>
  =\frac{1}{N_{\rm site}} \sum_{a,b}\int\! d^3 \x \, [ k_j (u^\alpha)^*_c(\x-\mathbf{R}_a)  u^{i \alpha}_{v}(\x-\mathbf{R}_b) -i(u^\alpha)^*_c(\x-\mathbf{R}_a) \partial_j u^{i \alpha}_{v}(\x-\mathbf{R}_b) ],
\end{equation}
The first term vanishes due to the orthogonality of LCAO. The second term can be written as 
\begin{equation}
 	\left< \psi^\alpha_\k|  p_j|\chi_{i\alpha,\k} \right>=-i\sum_{a,b}\int_{\text{unit cell}} d^3 \x \, (u^\alpha)^*_c(\x-\mathbf{R}_a) \partial_j u^{i \alpha}_{v}(\x-\mathbf{R}_b).
\end{equation}
 Due to the angular momentum conservation, we have 
\begin{equation}
	\left< \psi^\alpha_\k|  p_j|\chi_{i\alpha,\k} \right>=-i\delta_{ij}\sum_{a,b}\int_{\text{unit cell}} d^3 \x \, (u^\alpha)^*_c(\x-\mathbf{R}_a)  \partial_i u^{i \alpha}_{v}(\x-\mathbf{R}_b).
\end{equation}
Following Ref.~\cite{Haus2016}, we obtain 
\begin{equation}
\left< \psi^\alpha_\k|  p_i|\chi_{i\alpha,\k} \right>= |\hat{e}\cdot \vec{p}_{\rm cv}|,
\end{equation}
where $\hat{e}$ is any unit vector and $p^x_{\rm cv}=p^y_{\rm cv}=p^z_{\rm cv}$ with the definition
\begin{equation}
	p^i_{\rm cv}=-i\sum_{a,b}\int_{\text{unit cell}} d^3 \x \, (u^\alpha)^*_c(\x-\mathbf{R}_a) \partial_i u^{i \alpha}_{v}(\x-\mathbf{R}_b) .
\end{equation}
We also have the relation between $|\hat{e}\cdot \vec{p}_{\rm cv}|$ and
the parameter $E_p$ often use in the photonics literature~\cite{Haus2016}
\begin{equation}
|\hat{e}\cdot \vec{p}_{\rm cv}|^2=\frac{E_p m}{2}\,. 
\end{equation}
We then obtain 
\begin{equation}
	P^*=-\frac{i}{m \omega_{ph}}\sqrt{\frac{E_p m}{2}}\,.
\end{equation}
Note that the coupling depends on the frequency of photon. However, if we rewrite the coupling with electric field to coupling with gauge potential, there will be no dependence on the photon frequency. We can rewrite the coupling with photon \eqref{eq:photon} as 
\begin{equation}
	\label{eq:photon2}
	\mathcal{L}_{i} =
e\sqrt{\frac{E_p }{2 m}} (\psi_\alpha^\dagger \chi^\alpha_i A_i+\chi^\dagger_{i\alpha}\psi^\alpha A_i^*).
\end{equation}

\section{How to project to a Landau level}
\label{sec:project}
In this section, we provide the detail calculation for kinetic part of stress tensor in a specific Landau level. The electron field can be expanded in orbitals
\begin{equation}
	\psi(\x) = \sum_{Mm} \< \x | Mm\> c_{Mm},
\end{equation}
where $M$ labels the Landau level, and $m$ labels the states within the
Landau level.
In order to obtain the kinetic part of stress tensor, we need  to compute
\begin{equation}
	\int\!d\x\, W^\+(\x) D_z^2 \psi(\x) =
	-\!\int\!d\x\, d\x'\, \psi^\+(\x) \psi^\+(\x') V(\x-\x') \psi(\x') D_z^2 \psi(\x).
\end{equation}
Inserting the expansion over modes, and limiting to the $N$th Landau
level, this becomes
\begin{equation}
	- \int\!d\x\, d\x'
	\sum_{mnm'n'} \<Nm|\x\> \<Nm'|\x'\> V(\x-\x') \<\x'|Nn'\> D_z^2\<\x|Nn\> c^\+_{Nm} c^\+_{Nm'} c_{Nn'} c_{Nn} .
\end{equation}
Introducing the Fourier transform of the potential
\begin{equation}
\label{eq:V}
	V(\x-\x') = \int\!\frac{d\q}{(2\pi)^2}\, e^{i\q\cdot(\x-\x')} V(\q),
\end{equation}
the expression becomes
\begin{equation}
	-\! \sum_{mnm'n'} \int_\q V(\q) \!\int\!d\x\,\<Nm|\x\> e^{i\q\cdot\x} D_z^2 \<\x|Nn\>
	\!\int\!d\x'\,
	\<Nm'|\x'\> e^{-i\q\cdot\x'} \<\x'| Nn'\>
	c^\+_m c^\+_{m'} c_{n'} c_{n} . 
\end{equation}

Now we have
\begin{equation}
	\int\!d\x'\, \<Nm'|\x'\> e^{-i\q\cdot\x'} \<\x'| Nn'\>
	=   \< Nm' | e^{-i\q\cdot \hat\x} |Nn'\>
	= \< N | e^{-i\q\cdot \tilde{\mathbf{r}}} |N\> \<m'| e^{-i\q \cdot\R} |n'\> ,
\end{equation}
but
\begin{multline}
	\< N | e^{-i\q\cdot \tilde{\mathbf{r}}} |N\>
	= \< N| \exp\left[ -\frac {i \ell_B}{\sqrt2}
	(q^z a^\+ + q^{\bar z} a) \right] | N\>\\
	= e^{-x_q/2} \< N | \exp\left(-\frac {i \ell_B}{\sqrt2} q^z a^\+ \right)
	\exp\left(-\frac {i \ell_B}{\sqrt2} q^{\bar z} a \right) |N\>
	= e^{-x_q/2} L_N(x_q), \quad x_q = \frac{q^2\ell_B^2}2 \,,
\end{multline}
therefore
\begin{equation}
	\sum_{m'n'}\!\int\!d\x'\,
	\<Nm'|\x'\> e^{-i\q\cdot\x'} \<\x'| Nn'\> c^\+_{m'} c_{n'}
	= e^{-x_q/2} L_N (x_q ) \bar\rho(\q) .
\end{equation}

Analogously
\begin{multline}
	\int\!d\x\, \<Nm|\x\> e^{i\q\cdot\x} D_z^2\<\x| Nn'\>
	=   \frac{1}{2 } \< Nm | e^{i\q\cdot \hat\x} \frac{a^{\+2}}{\ell_B^2}|Nn\>\\
	= \frac{1}{2 \ell_B^2} \sqrt{(N+1)(N+2)}
	\< N | e^{i\q\cdot \tilde{\mathbf{r}}} |N+2\> \<m| e^{i\q \cdot\R} |n\>
	= - q_z^2 e^{-x_q/2} L_N^2 (x_q)
	\<m| e^{i\q \cdot\R} |n\>,
\end{multline}
therefore
\begin{equation}
	\sum_{mn}\!\int\!d\x\,
	\<Nm|\x\> e^{-i\q\cdot\x} D_z^2 \<\x| Nn\> c^\+_{m} c_{n}
	= -q_z^2 e^{-x_q/2} L_N^2(x_q) \bar\rho(-\q) .
\end{equation}
Finally we obtain
\begin{equation}
	\label{eq:T1}
	\int\!d\x\, W^\+(\x) D_z^2 \psi(\x)
	=  \int\!\frac{d\q}{(2\pi)^2}\, q_z^2 e^{-x_q} L_N(x_q)
	L_N^2(x_q) V(q) \bar\rho(\q)\bar\rho(-\q) .
\end{equation}
Similarly, for $N\ge2$
\begin{equation}
	\label{eq:T2}
	\int\!d\x\, W^\+(\x) D_{\bar z}^2 \psi(\x)
	= \int\!\frac{d\q}{(2\pi)^2}\, q_{\bar z}^2 e^{-x_q} L_N(x_q)
	L_{N-2}^2(x_q) V(q) \bar\rho(\q) \bar\rho(-\q),
\end{equation}
while for $N=0$ or 1 the expression is obviously zero due to the presence
of two lowering operators $D_{\bar z}$ acting on $\psi$.
Equations \eqref{eq:T1} and \eqref{eq:T2} are used in Sec. \ref{sec:TLL} to obtain the explicit form of the kinetic part of stress tensor on a specific Landau level.

\section {$i\psi^\dagger \partial_t \psi$ and $T^{\rm kin}_{z \bar{z}}$ }
\label{sec:psidtpsi}

In this Appendix, we derive the explicit form of $i\psi^\dagger
\partial_t \psi$ in the lowest Landau level.  The field equation reads
\begin{equation}
  i\partial_t \psi(\mathbf{x})= - \frac{1}{m^*} \left(D_z D_{\bar z} +D_{\bar z}D_z\right)\psi(\mathbf{x})
  +\int\! d \mathbf{x}'\, V(\mathbf{x}-\mathbf{x}')\psi^{\dagger}(\mathbf{x}')\psi(\mathbf{x}')\psi(\mathbf{x}),
\end{equation}
we then use the constraint equation $2\ell_B^2 D_z D_{\bar z} \psi=-N\psi$ and the commutator \eqref{eq:comm} from that we get
\begin{equation}
	\label{eq:dt}
  i\!\int\! d \mathbf{x} \, \psi^\dagger \partial_t \psi
  =   I_0+(N+\frac{1}{2})\omega_c N_e,
\end{equation}
with $N_e$ being the total electron number in the conductance band and 
\begin{equation}
   I_0 = \int\! d \x \, d\x'\, V(\x-\x') \psi^\+(\x) \psi^\+(\x')\psi(\x')\psi(\x)  = \int_{\q}
	V(\q) e^{-q^2\ell_B^2/2} \bar\rho(\q) \bar\rho(-\q).
\end{equation}
The first term on the on the right hand side of Eq. \eqref{eq:dt} is
twice the interacting energy and second term is  the kinetic
energy of electrons\footnote{In Ref.~\cite{Nguyen2014LLL} we
eliminate the second term in the LLL case by introducing the Landr\'e factor
$\mathfrak{g}=2$ }.  We also have
\begin{align}
   \int\! d \x \, T^{\rm kin}_{z \bar z}&=\int\! d \x \, \frac{1}{2m^*} \left( D_z \psi^\dagger D_{\bar z} \psi+ D_{\bar z} \psi^\dagger D_{z} \psi \right) \nonumber\\
	&=\int\! d \x \, \frac{1}{2m^*} \left( -2  \psi^\dagger D_{z} D_{\bar z} \psi- \frac{eB}{2c}\psi^\dagger  \psi\right) \nonumber\\
	&=\frac{\omega_c}{2}(N+\frac{1}{2})\! \int\! d \x \, \rho=(N+\frac{1}{2})\frac{N_e \omega_c}{2}. 
\end{align}
Where we used the contraint $2\ell_B^2 D_z D_{\bar z} \psi=-N\psi$ to obtain the last equality.

\section{Two sum rules}
\label{sec:sumrule}

This Appendix is not directly related to Raman scattering, but contains
some exact sum rules. First we write down formulas for the full stress tensor.  For the model Hamiltonian, at the long wave length regime $\mathbf{k}\sim 0$, the full stress tensor is the same as $T^{\rm kin}$ \cite{Nguyen2014LLL}.
For the general case, one needs to take into account the potential-energy
term in the Lagrangian.  When the metric varies with time (but remains
uniform in space), the potential changes according to
\begin{equation}
  V(\x) \to V(\sqrt{g_{ij} x^i x^j}),
\end{equation}
and so the Fourier transform changes as
\begin{equation}
  V(\q) \to V(\sqrt{g^{ij} q_i q_j}) .
\end{equation}
Given that the stress tensor is give by $\delta S= \frac12 \int\!dx\, T_{ij}\delta
g_{ij}$, the potential part of the stress-energy tensor is then
\begin{equation}
  \int\!d\x\, T_{ij}^{\rm pot} = \frac12 \int_\q \frac{q_i q_j} q V'(q) \rho(\q)\rho(-\q) .
\end{equation}
This can be projected to the $N$th Landau level to become
\begin{equation}\label{Tpot}
  \int\!d\x\, T_{ij}^{\rm pot} =
  \frac12 \int_\q e^{-x_q}
  \left[L_N ( x_q )\right]^2
  \frac{q_i q_j} q V'(q)
  \bar\rho(\q) \bar\rho(-\q) .
\end{equation}
It is interesting to compare the formula to to that of the kinetic
stress tensor, Eq.~(\ref{Tkin-summarized}): the ``stretching
operator'' $(q_i q_j/q)\d_q$ now acts on the potential $V(q)$ instead
of acting on the form-factor.  The full stress tensor is the sum
of the kinetic stress tensor, Eqs.~(\ref{eq:Tzkin}), and the potential
stress tensor, Eq.~(\ref{Tpot}).  It can be written as
\begin{equation}
  \int\!d\x\, T_{ij}^{\text{full}} = \frac12 \sum_\q
  \frac{q_i q_j}q \frac\d{\d q} \left\{
  e^{-x_q} [L_N(x_q)]^2 V(q) \right\}
  \bar\rho(\q) \bar\rho(-\q) .
\end{equation}

%
%
We define the spectral densities \cite{Golkar}
\begin{align}
	\rho_T(\omega) &= \frac1N \sum_n |\< n| \int\!d\x\, T_{zz} |0\>|^2 \delta(\omega-E_n),\\
	\bar\rho_T(\omega) &= \frac1N \sum_n |\< n| \int\!d\x\, T_{\bar z\bar z} |0\>|^2 \delta(\omega-E_n),
\end{align}
where $N$ is the total number of electrons, $|0\>$ is the ground
state, the sum is taken over all excited states $|n\>$ in the lowest Landau level, and $E_n$ is
the energy of the state $|n\>$. The two spectrum densities satisfy the sum rules~\cite{Golkar}
\begin{align}
	\int_0^\infty \frac{d\omega}{\omega^2} \left[
	  \rho_T(\omega) - \bar\rho_T(\omega) \right] &=
        \frac{\bar s}4 \,,
        \\
	\int_0^\infty \frac{d\omega}{\omega^2} \left[
	\rho_T(\omega) + \bar\rho_T(\omega) \right] &= S_4,
\end{align}
where $\bar s$ is the ``guiding center spin''~\cite{Haldane:2011},
which is equal to $(\mathcal{S}-1)/2$ on the LLL where $\mathcal S$ is
the shift, and $S_4$ is the coefficient in front of $(k\ell_B)^4$
in the static structure factor.

\bibliography{Rm}

\end{document}